# Fractional optical vortices in a uniaxial crystal

T A Fadeyeva[1], A F Rubass[1], I S Valkov[2] and A V Volyar[1]

[1] General Physics Department, Taurida National University, Simferopol, Ukraine

[2] John Atanasoff Technical College, Plovdiv, Bulgaria

e-mail: volyar@crimea.edu

**Abstract** We have analyzed the solutions to the vector paraxial wave equation in the unbounded uniaxial crystal in the form of the transverse electric (TE) and transverse magnetic (TM) mode beams transporting the fractional optical vortices in the circularly polarized components. We revealed that the TE and TM beams have asymmetric structure in distributing the local elliptic polarization over the beam cross-section and form two sets of singular beams depending on the real or imaginary value of the free K parameter. We found that the fractional optical vortex born in the left handed circularly polarized component of the beam with the real K parameter can exist in the form of the holistic structure within a small crystal lengths much smaller the Rayleigh while the beams with the imaginary K-parameter cannot maintain fractional optical vortices in free state. However, such beam types can generate the singly charged optical vortices in far field. We also revealed that the energy efficiency and spin-orbit coupling are defined by the angular spectrum of the beam. The beam with the real K parameter is characterized by broad spectrum of plane waves propagating at small angles to the crystal optical axis. The beams with the imaginary K parameter are shaped by two conical fans of plane waves. It is this circumstance that define a very high value of the energy efficiency.

**Keywords:** optical vortices, uniaxial crystals, fractional charge

## I. Introduction

The problem of the vortex beams bearing optical vortices with fractional topological charges is one of the most intriguing parts of the advanced singular optics. The first experiments with fractional vortex beams on the base of computer-generated holograms considered by Soskin et. al [1,2] as far back as in the beginning of $90^{th}$ showed the principal instability of such wave constructions when propagating. Later the theoretical analysis of the Gaussian beam diffracted by the spiral plate with a fractional phase step presented by Berry [3] have shown that the diffracted beam splits at once into the infinite series of the standard vortex beams with integer-order topological charged evolving in space by a complex way. The detailed experimental preparation of the phase structure of the fractional vortex beams of different kinds with help of the holographic technique researched by Padgett et al [4] have corroborated such a theoretical proposition. Subsequent study of the conical beams [5, 6] has also shown that fractional optical vortices cannot exist in free state even in the initial plane but are of the complex array of singly charged optical vortices. The vector features of the fractional vortex beams in free space and homogeneous isotropic media were considered in the paper [7].

At the same time, there is a number of mechanical and quantum mechanical analogues of the phase defects with a fractional phase step, e.g. in the Ahoronov –Bohm effect and in the surface waves in liquid [8], pointing out a possibility to observe the fractional vortex beam in free state in optics. Recently we have shown [9] that there is at least one types of vortex beams with the ½-topological charge (the so-called erf-G beams) that can contain the fractional optical vortex in free state near the initial plane. However, such a vortex state is a very unstable one splitting into the integer-order vortex array when propagating within a Rayleigh length. Also it was not as yet developed the reliable technique for generating such wave constructions. However, there is the well-known technique for creating the integer order vortices of highest orders on the base of uniaxial [10-16], biaxial [17-20] and biaxially-induced crystals [21-24]. Thus, it is of a great theoretical and practical interest to bring to light the basic physical processes responsible for the nucleation and propagation of the fractional optical vortices in light beams traveling through uniaxial crystals as the potential media for creating and manipulating such wave constructions.

The aim of our paper is to analyze the diffractive processes of the paraxial optical beams in a uniaxial crystal resulting in the generation of fractional optical vortices and to estimate their energy efficiency.

**II Asymmetric structure of TE and TM mode beams**

We consider the propagation of the monochromatic paraxial singular beams with fractional optical vortices along the optical axis of the unbounded uniaxial crystal with the permeability tensor $\hat{\varepsilon} = diag\{\varepsilon_{jj}\}$, $\varepsilon_{11} = \varepsilon_{22} = n_1^2, \varepsilon_{33} = n_3^2$, where $n_1 = n_2 = n_o$ and $n_3$ are the refractive indices along the tensor principle axes, respectively. The electric field of the paraxial beam can be treated as $\tilde{\mathbf{E}}(x,y,z) = \mathbf{E}(x,y,z)\exp(ik_o z)$, where $\mathbf{E}(x,y,z)$ stands for the complex amplitude of the field, $k_o = n_o k$, $k$ is the wavenumber. Then the vector paraxial wave equation for the transverse field components $\mathbf{E}_\perp = \mathbf{e}_x E_x + \mathbf{e}_y E_y$ is written in the form [12]:

$$\left(\nabla_\perp^2 + 2ik_o \partial_z\right)\mathbf{E}_\perp = \alpha \nabla_\perp \left(\nabla_\perp \mathbf{E}_\perp\right), \qquad (1)$$

whereas the longitudinal component can be found as

$$E_z \approx \frac{i n_o}{k n_e^2} \nabla_\perp \mathbf{E}_\perp, \qquad (2)$$

where $\nabla_\perp = \mathbf{e}_x \partial_x + \mathbf{e}_y \partial_y$, $\alpha = \Delta\varepsilon / \varepsilon_{11}$, $\Delta\varepsilon = \varepsilon_{11} - \varepsilon_{33}$.

When dealing with the vortex-beams it is convenient to use the complex variables [12, 15]

$$u = x + i y, \quad v = x - i y, \qquad (3)$$

so that

$$\partial_u = \partial_x - i\partial_y = \frac{e^{-i\varphi}}{2}\left(\partial_r - \frac{i}{r}\partial_\varphi\right), \qquad (4)$$

$$\partial_v = \partial_x + i\partial_y = \frac{e^{i\varphi}}{2}\left(\partial_r + \frac{i}{r}\partial_\varphi\right), \qquad (5)$$

where $(r,\varphi)$ are the polar coordinates. The operators $\partial_u$ and $\partial_v$ can be treated as the operators of the annihilation and birth of optical vortices, respectively, while

$$\nabla_\perp^2 \equiv 4\partial_{uv}^2, \qquad \nabla_\perp \mathbf{E}_\perp = \partial_v E_+ + \partial_u E_-, \qquad (6)$$

In the last equation (6), the transverse electric field is represented in the circularly polarized basis:

$$E_+ = E_x - iE_y, \; E_- = E_x + iE_y. \qquad (7)$$

Now the vector wave equation can be rewritten for each field component as

$$\left(\nabla_\perp^2 + 2ik_o\partial_z\right)E_+ = 2\alpha\partial_u\left(\partial_v E_+ + \partial_u E_-\right), \qquad (8)$$

$$\left(\nabla_\perp^2 + 2ik_o\partial_z\right)E_+ = 2\alpha\partial_v\left(\partial_v E_+ + \partial_u E_-\right). \qquad (9)$$

The simplest solutions of the equations system (8) and (9) have the form of the transverse electric TE and transverse magnetic TM mode beams:

*the TE mode beams,* $E_z = 0$

$$E_+^{(o)} = \partial_u \Psi_o, \; E_-^{(o)} = -\partial_v \Psi_o, \qquad (10)$$

while the generatrix function $\Psi_o$ obeys the paraxial wave equation

$$\left(\nabla_\perp^2 + 2ik_o\partial_z\right)\Psi_o = 0, \qquad (11)$$

*the TM mode beams,* $H_z = 0$

$$E_+^{(o)} = \partial_u \Psi_e, \; E_-^{(o)} = \partial_v \Psi_e, \qquad (12)$$

while the generatrix function $\Psi_e$ obeys the paraxial wave equation

$$\left(\nabla_\perp^2 + 2i\frac{k_o}{1-\alpha}\partial_z\right)\Psi_e = 0, \qquad (13)$$

From whence the TE mode beams are characterized by the ordinary refractive index $n_o$ and the wavenumber $k_o$ while the TM mode beams are associated with the extraordinary refractive index $n_e = \frac{n_o}{1-\alpha} = \frac{n_3^2}{n_o}$ and the wavenumber $k_e = n_e k$.

Thus, the solutions of the vector equation (1) are held back by the solutions of the scalar equations (11) and (13). We will find the wave functions $\Psi_o$ and $\Psi_e$ in the form

$$\Psi_o^{(q)} = N_o(z) F_q^o(U_o, V_o) G_o(u, v, z), \qquad (14)$$

$$\Psi_e^{(q)} = N_e(z) F_q^e(U_e, V_e) G_e(u, v, z), \qquad (15)$$

where $U_{o,e} = \dfrac{u}{w_0 \sigma_{o,e}}$, $V_{o,e} = \dfrac{v}{w_0 \sigma_{o,e}}$, $N_{o,e}(z)$ is the normalizing coefficients, $\sigma_{o,e} = 1 + i z / z_{o,e}$, $z_{o,e} = k_{o,e} w_0^2 / 2$, $w_0$ is the radius of the beam waist at $z = 0$, $G_{o,e} = \exp(-uv / w_0^2 \sigma_{o,e}) / \sigma_{0,e}$ are the Gaussian envelopes of the ordinary and extraordinary beams, while the modulation functions $F_{o,e}^{(q)}$ obey the Helmholtz-Kiselev equation [25]:

$$\left(4 \frac{\partial^2}{\partial U_{o,e} \partial V_{o,e}} + K_{o,e}^2\right) F_{o,e}^{(q)} = 0, \tag{16}$$

where the parameter $K_{o,e}$ can be arbitrary complex value and $q$ is some number.

We choose the solutions of the eqs. (16) in the form of the erf-G beams (*error function-Gaussian beams*) with $q = 1/2$:

$$F_{1/2}^o = \int_0^{2\pi} e^{i\frac{\phi}{2}} \exp\left\{-\frac{K_\perp^o r}{\sigma_o} \cos(\phi - \varphi)\right\} d\phi = \int_0^{2\pi} e^{i\frac{\phi}{2}} \exp\left\{-\frac{K_\perp^o w_0}{2}\left[U_o e^{-i\phi} + V_o e^{i\phi}\right]\right\} d\phi =$$

$$= -\frac{2i\sqrt{\pi} e^{i\varphi/2}}{\Re_o} \left\{ e^{-\Re_o^2/2} \mathrm{erf}\left(i \Re_o \sin\frac{\varphi}{2}\right) + e^{\Re_o^2/2} \mathrm{erf}\left(\Re_o \cos\frac{\varphi}{2}\right) \right\}, \tag{17}$$

$$F_{1/2}^e = \int_0^{2\pi} e^{i\frac{\phi}{2}} \exp\left\{-\frac{K_\perp^e r}{\sigma_e} \cos(\phi - \varphi)\right\} d\phi = \int_0^{2\pi} e^{i\frac{\phi}{2}} \exp\left\{-\frac{K_\perp^e w_0}{2}\left[U_e e^{-i\phi} + V_e e^{i\phi}\right]\right\} d\phi =$$

$$= -\frac{2i\sqrt{\pi} e^{i\varphi/2}}{\Re_e} \left\{ e^{-\Re_e^2/2} \mathrm{erf}\left(i \Re_e \sin\frac{\varphi}{2}\right) + e^{\Re_e^2/2} \mathrm{erf}\left(\Re_e \cos\frac{\varphi}{2}\right) \right\}, \tag{18}$$

where $\Re_{o,e} = \sqrt{2 K_\perp^{(o,e)} r / \sigma_{o,e}}$, $K_\perp^{(o,e)} = i K_{o,e} / w_0$, $\mathrm{erf}(x)$ stands for the error function, and $N_{o,e} = \exp\left\{-\left(K_\perp^{(o,e)} w_0\right)^2 / 4 \sigma_{o,e}\right\}$.

Thus, the solutions of the vector wave equation (1) can be obtained for the *TE mode beams* in the form

$$E_+^o = \partial_u \Psi_{1/2}^o = -\frac{N_o G_o}{\sigma_o} \left\{ F_{-1/2}^o + \frac{r e^{-i\varphi}}{r_\perp^o \sigma_o} F_{1/2}^o \right\}, \tag{19}$$

$$E_-^o = -\partial_v \Psi_{1/2}^o = \frac{N_o G_o}{\sigma_o} \left\{ F_{3/2}^o + \frac{r e^{i\varphi}}{r_\perp^o \sigma_o} F_{1/2}^o \right\}, \tag{20}$$

and for the *TM mode beams* in the form

$$E_+^e = \partial_u \Psi_{1/2}^e = -\frac{N_e G_e}{\sigma_e} \left\{ F_{-1/2}^e + \frac{r e^{-i\varphi}}{r_\perp^e} F_{1/2}^e \right\}, \tag{21}$$

$$E_-^e = \partial_v \Psi_{1/2}^e = -\frac{N_e G_e}{\sigma_e} \left\{ F_{3/2}^e + \frac{r e^{i\varphi}}{r_\perp^e} F_{1/2}^e \right\}, \tag{22}$$

where $r_\perp^{(o,e)} = K_\perp^{(o,e)} w_0^2 / 2$ and we made use of the relations

$$\partial_{U_{o,e}} F_q^{(o,e)} = -\left(K_\perp^{(o,e)} w_0/2\right) F_{q-1}^{(o,e)}, \quad \partial_{V_{o,e}} F_q^{(o,e)} = -\left(K_\perp^{(o,e)} w_0/2\right) F_{q+1}^{(o,e)}, \tag{23}$$

$$F_q^{(o,e)} = \int_0^{2\pi} e^{iq\phi} \exp\left\{-\frac{K_\perp^{\{o,e\}} r}{\sigma_{o,e}} \cos(\phi-\varphi)\right\} d\phi. \tag{24}$$

Besides, the functions $F_q^{(o,e)}$ in eqs (19) - (22) are expressed in the explicit form as

$$F_{-1/2}^{(o,e)} = -\frac{i2\sqrt{\pi} e^{-i\varphi/2}}{\Re_{o,e}} \left\{ e^{-\Re_{o,e}^2/2} erf\left(i\Re_{o,e} \sin\frac{\varphi}{2}\right) - e^{\Re_{o,e}^2/2} erf\left(\Re_{o,e} \cos\frac{\varphi}{2}\right)\right\}, \tag{25}$$

$$F_{3/2}^{(o,e)} = -\frac{e^{i3\varphi/2}}{\Re_{o,e}} \left\{ e^{-\Re_{o,e}^2/2} \left[ 2i\sqrt{\pi}\left(1+\frac{2}{\Re_{o,e}^2}\right) erf\left(i\Re_{o,e}\sin\frac{\varphi}{2}\right) + \frac{8}{\Re_{o,e}}\sin\frac{\varphi}{2} e^{\Re_{o,e}^2 \sin^2\frac{\varphi}{2}} \right] + \right.$$

$$\left. +ie^{\Re_{o,e}^2/2}\left[ 2\sqrt{\pi}\left(1-\frac{2}{\Re_{o,e}^2}\right) erf\left(\Re_{o,e}\cos\frac{\varphi}{2}\right) + \frac{8}{\Re_{o,e}}\cos\frac{\varphi}{2} e^{-\Re_{o,e}^2 \cos^2\frac{\varphi}{2}} \right]\right\}. \tag{26}$$

For more general case of the highest order TE and TM beams we obtain from eqs (19)-(22)

$$\left(E_+^{(m,n)}\right)_{o,e} = \partial_v^n \partial_u^{m+1} \Psi_{1/2}^{(o,e)}, \quad \left(E_-^{(m,n)}\right)_{o,e} = s_{o,e} \partial_v^{n+1} \partial_u^m \Psi_{1/2}^{(o,e)}, \tag{27}$$

where $s_o = 1$, $s_e = -1$.

It should be noted that the fields (19)-(22) of the TE and TM modes have asymmetric structure and are not linearly polarized in each point of the cross-section [9] in contrast to the standard TE and TM modes beams [12]. Besides, their field structure is transformed when propagating. Nevertheless, the corresponding longitudinal components of the electric or magnetic fields are strongly zero.

We will consider the case when only $E_+$ component exist at the initial plane $z=0$, i.e $E_-(z=0)=0$. Then the field is described by the expression:

$$\mathbf{E}(r,\varphi,z) = \mathbf{E}_o(r,\varphi,z) + \mathbf{E}_e(r,\varphi,z), \tag{28}$$

so that $K_\perp^o = K_\perp^e = K_\perp$, $r_\perp^o = r_\perp^e = r_\perp$ and $\mathbf{E}(r,\varphi,z=0) = \mathbf{e}_+ 2E_+^o(r,\varphi,z=0)$. The corresponding intensity and phase distributions for the case $K_\perp$ being the *real value* is shown in Figure 1 and Figure 2, respectively. The characteristic feature of the phase distribution is *the holistic helix-like surface* with the step $\Delta\Phi = \pi$ (Figure 2) in contrast to that for the case where $K_\perp$ is the *imaginary value* (see Figure 3 a, b) with the standard representation of the fractional vortex in the form of the infinite series of the integer order optical vortices (see e.g. [2,3]).

Undivided attention should be drawn to the black line in the form of the broken circle cutting the intensity spike in Figure1. In order to analyze such a wave structure let us write the condition of the zeros in the field (19) $E_o(r,\varphi,z=0)=0$ *far from the axis* $r=0$ when $z=0$:

$$\left(1+\frac{r}{r_\perp}\right) e^{-K_\perp r} erf\left(i\sqrt{2K_\perp r}\sin\frac{\varphi}{2}\right) + \left(1-\frac{r}{r_\perp}\right) e^{K_\perp r} erf\left(\sqrt{2K_\perp r}\cos\frac{\varphi}{2}\right) = 0. \tag{29}$$

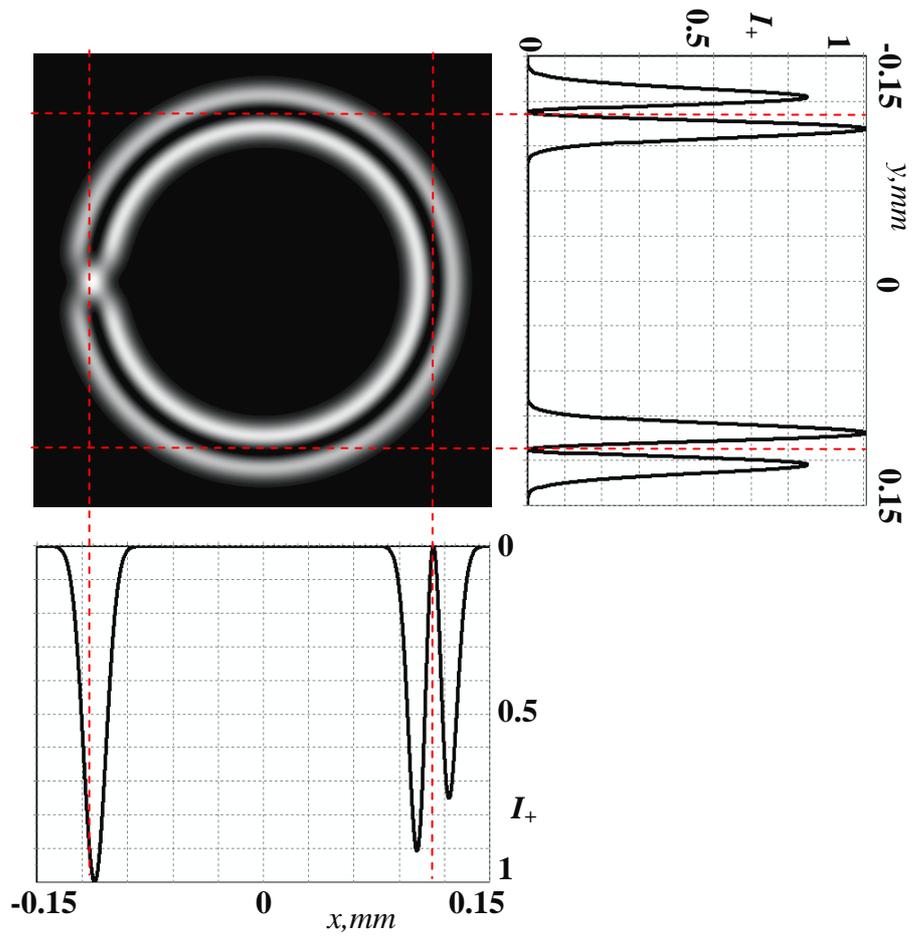

Fig.1 Intensity distribution $E_+$ component of the TE mode at: $z = 0$ $w_0 = 30\,\mu m$, $K_\perp = 3\cdot 10^5\,m^{-1}$

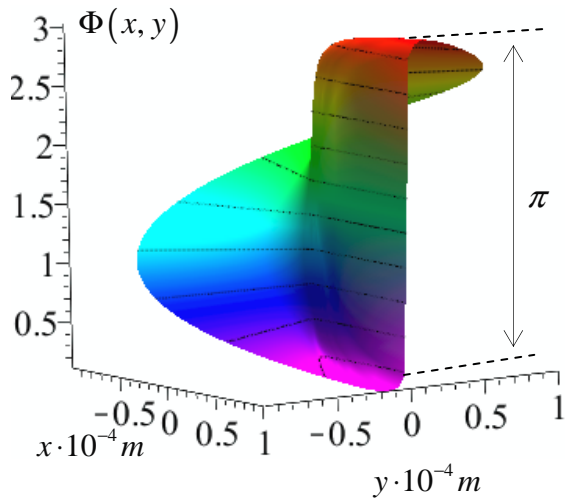

Fig.2 Phase distribution $\Phi(x,y)$ near the axis $r=0$ in the $E_+$ component at the initial plane $z=0$

First of all, the second term in eq. (29) is much greater than the first one with the exception of the region near $\varphi = \pi$ because of $\exp(K_\perp r_\perp) \approx 7,21\cdot 10^{48}$ and $\exp(-K_\perp r_\perp) \approx 1.4\cdot 10^{-49}$ for $K_\perp = 5\cdot 10^5\,m^{-1}$, $w_0 = 30\,\mu m$ but it vanishes at the circle

$$r = r_\perp = \frac{K_\perp w_0^2}{2}. \qquad (30)$$

On the other hand, the first term in eq. (29) is not the exact zero along the circle (30) excepting for the point $\varphi = 0$. Near the point $\varphi = \pi$ and $K_\perp r \gg 1$ its asymptotic value can be described by the expression

$$\left(1 + \frac{r}{r_\perp}\right) e^{-K_\perp r} erf\left(i\sqrt{2K_\perp r}\sin\frac{\varphi}{2}\right) \sim i\left(1 + \frac{r}{r_\perp}\right) \frac{e^{-K_\perp r}}{\sqrt{\pi}\sqrt{2K_\perp r}} \exp\left(2K_\perp r \sin^2\frac{\varphi}{2}\right), \qquad (31)$$

forming the bright spot at the end of the black ring $r = r_\perp$ in a Figure1 while near the point $\varphi = 0$ it can be approximated as

$$\left(1 + \frac{r}{r_\perp}\right) e^{-K_\perp r} erf\left(i\sqrt{2K_\perp r}\sin\frac{\varphi}{2}\right) \sim i\left(1 + \frac{r}{r_\perp}\right) \frac{2\,e^{-K_\perp r}}{\sqrt{\pi}} \sqrt{2K_\perp r}\sin\frac{\varphi}{2}. \qquad (32)$$

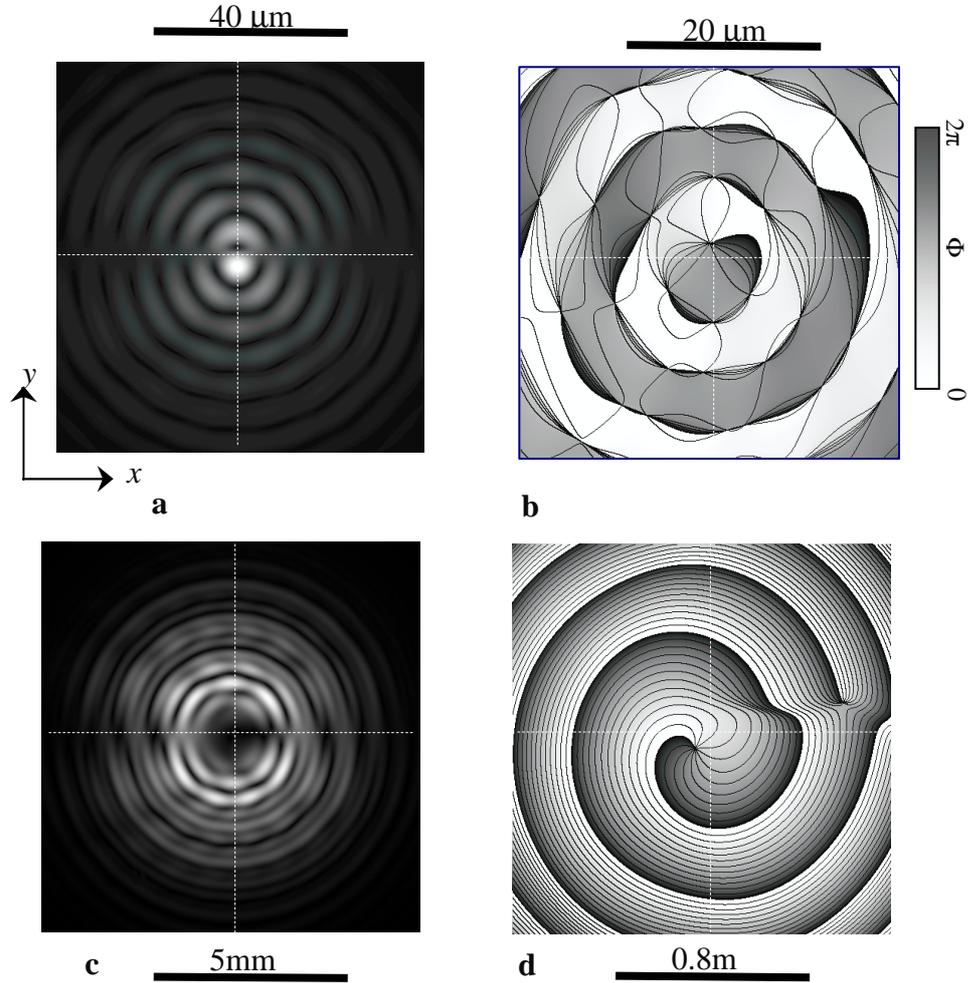

Fig.3 Intensity (a, c) and phase (b, d) distributions in TE mode with $w_0 = 30\,\mu m$, for (a, b) the $E_+$ component, $K_\perp = i5.5 \cdot 10^5 m^{-1}$ at the initial plane $z = 0$ and for (c, d) the $E_-$ component, $K_\perp = 5.5 \cdot 10^5 m^{-1}$ at the initial plane $z = 0.8m$

It means that the first term is imaginary one over all region $\varphi = (0, 2\pi)$ while the second term is real. Thus, the real and imaginary parts in eq. (29) vanish simultaneously at the point $\varphi = 0, r = r_\perp$.

Together with the factor $\exp(-i\varphi/2)$ they form very intricate of the phase singularity whose phase portrait in vicinity of the circle (30) is illustrated by Figure 4.

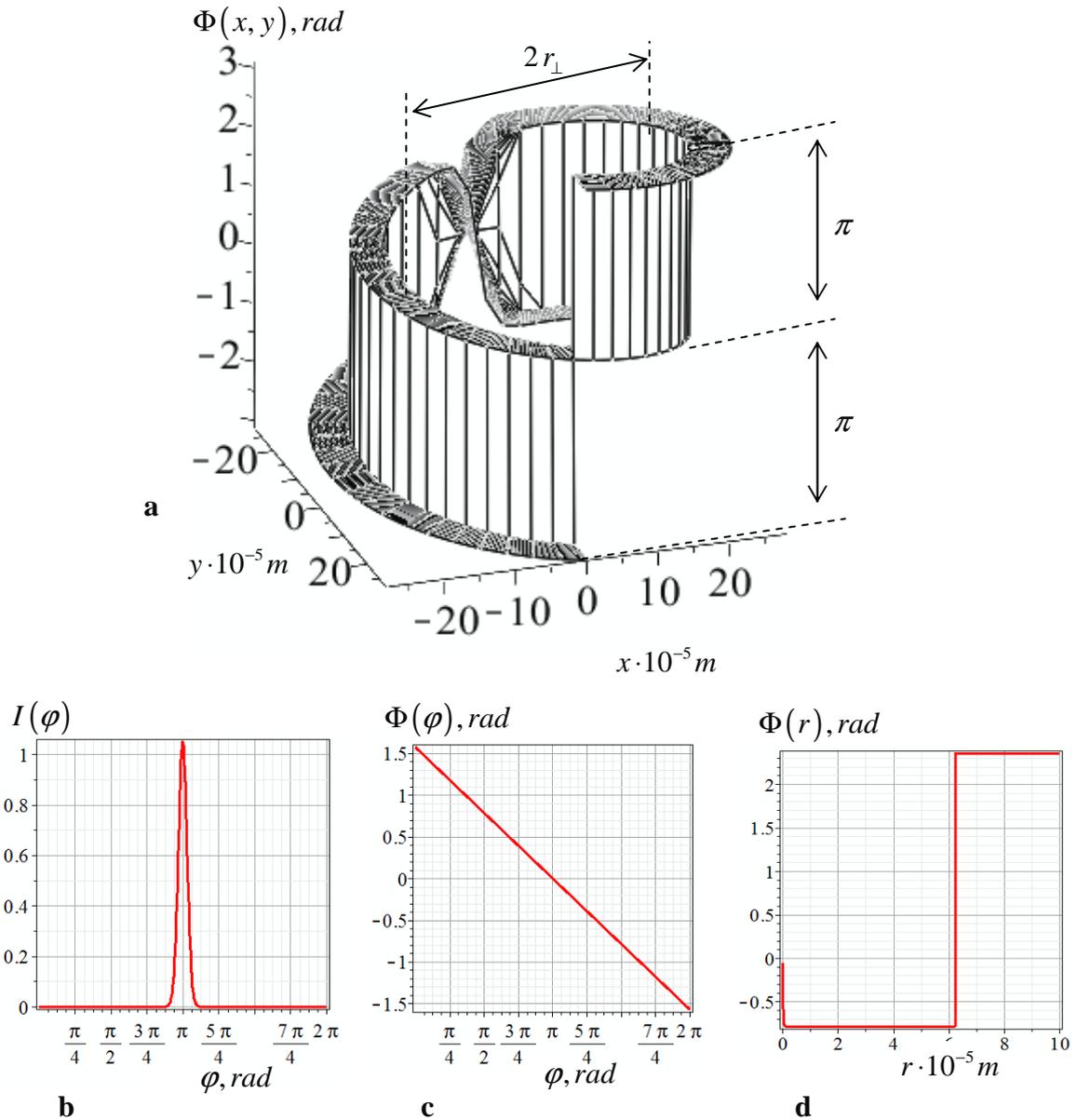

Fig.4 Phase distribution $\Phi(x,y)$ (a) and variations of intensity $I(\varphi)$ (b), phase $\Phi(\varphi)$ (c) along the circle of the radius $r_\perp = K_\perp w_0^2 / 2$ and also (d) phase variations $\Phi(r)$ along the ray $\varphi = \pi/2$

### III. The evolution of field states along the crystal

The inner structure of the centered and ring phase singularities described above is broken down when propagating the beam. We can for convenience outline two counter processes: 1) the beam diffraction and 2) spin-orbit coupling. The spin-orbit coupling result in nucleating something like a holistic screw dislocation (an optical vortex) with a fractional phase step proportional to $q = 2 - 1/2$ [15,26] in the vector beam with $K_\perp$ real while the diffraction process splinters it into a set of integer order optical vortices [9]. However, the process of the fractional vortex decay is not still instantaneous one. This displays itself little by little when propagating the beam as it shown in Figure 5.

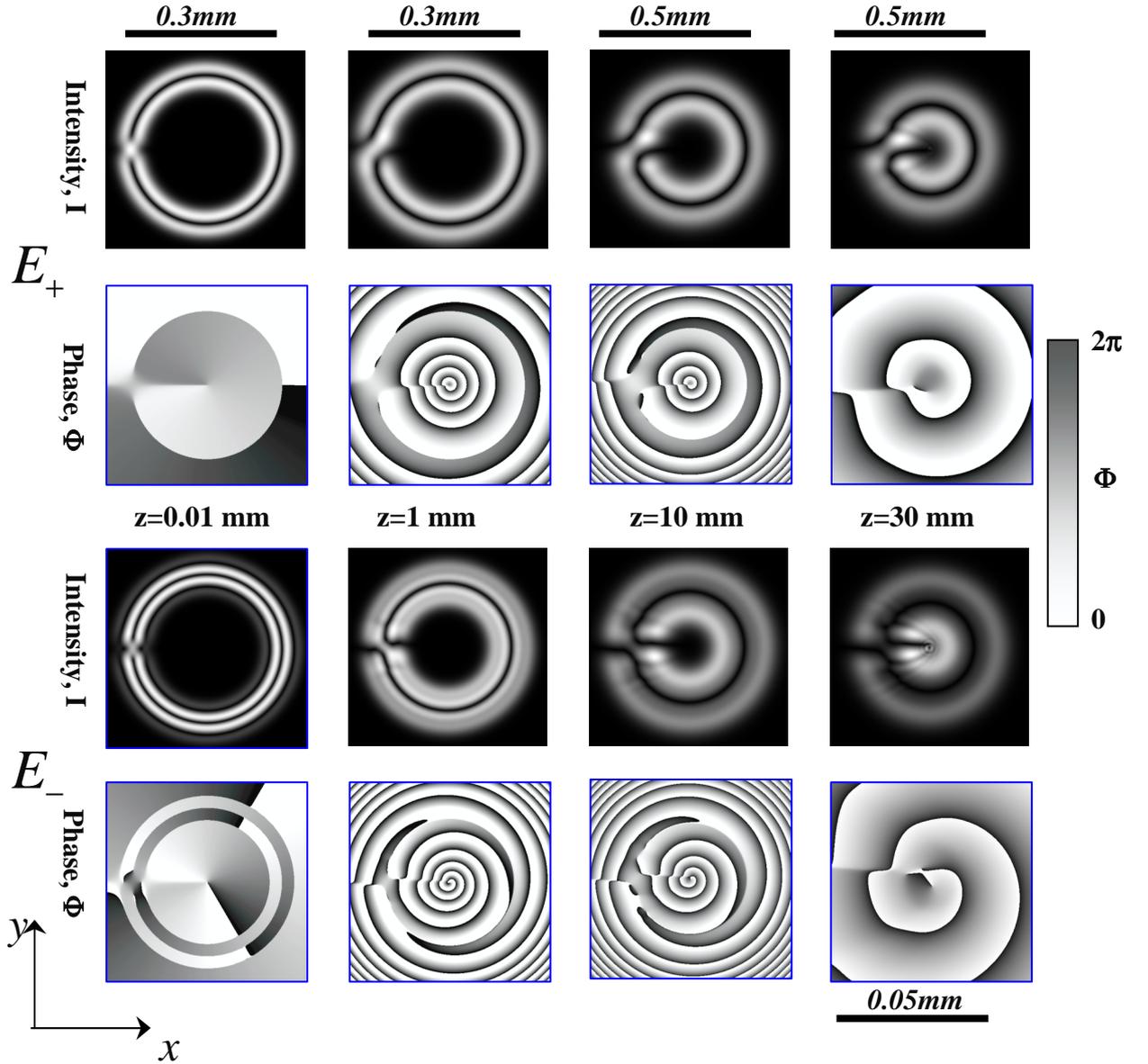

Fig.5 Intensity I and phase $\Phi$ distributions in the $E_+$ and $E_-$ components of TE mode beams of the lowest order along the length $z$ of the $LiNbO_3$ crystal $n_1 = 2.3$, $n_3 = 2.2$, $w_0 = 30\,\mu m$, $K_\perp = 3 \cdot 10^5 \, m^{-1}$

The non-disturbed state of the fractional vortex with $q = -1/2$ in the $E_+$ component and with $q = 3/2$ in the $E_-$ component keeps near the axis up to the $z=1mm$ crystal length. At the same time there is nucleation of integer order optical vortices at the periphery, in particular near the black broken ring in both components. At the length about $z=1cm$ we observe at last the nucleation of the vortex dipole near the axis in the $E_+$ component and the ensemble of three vortices (two with $q=+1$ and one with $q=-1$) in the $E_-$ component. At far field the vortex patterns in both component are utterly complicated coming to be like that shown in Figure 3 a,b for the $E_+$ component and in Figure 3c,d for the $E_-$ component. It is interesting to note that there is not the double-charged vortex at the axis as it could be expected [12]. Instead of it, we observe two separated singly charged optical vortices with the same sign of the topological charges (see Figure 3 d). Moreover, the energy exchange between circularly polarized components runs very slowly. The field transformations caused by the diffraction processes are more intense than those associated with the spin-orbit coupling. The vector structure of the beam changes very slowly.

Absolutely other situation occurs in the vortex beam with the imaginary $K_\perp$ parameter. At the initial plane *z=0* we observe a complex array of the integer order optical vortices (Figure 3 a,b). The characteristic feature of each TE and TM modes with fractional optical vortices in their circularly polarized components is of the inherent asymmetry of the field distribution [9] that defines the further evolution of the field. The diffraction processes rebuild very quickly the field structure: the speckle-like intensity distribution turns into the regular C-like field without the black broken ring at the crystal distance about 1 – 2 mm (see Figure 6). At the same time, the intense energy exchange between circularly polarized components results in the periodical conversion of the right and left-handed circular polarizations located at the intensity maximum of the C-like pattern. The uniform circular polarized field changes into the space variant polarization field with local linear polarization at the distance z=2.5mm. Then the field transforms into one with the dominated left hand circular polarization etc.

It is interesting to compare the vortex structures near the optical axis in the left handed circularly polarized components in the beams with the real and imaginary $K_\perp$ parameters (see Figure 7) at far diffraction field. The set of the equi-phase lines in Figure 7a for the real $K_\perp$ parameter has a smooth shape without any phase steps. It means that the left-handed field component does not carry over any centered optical vortices. At the same time the $E_-$ component in the beam with the imaginary $K_\perp$ parameter has the single optical vortex with a positive topological charge $l=1$ (Figure 7b) in contrast to that of the initial standard beams [12] bearing the double charged optical vortices near the crystal optical axis in the $E_-$ component. Moreover, there are not any integer-charged vortices at the periphery too safe for a weak phase perturbation in both circularly polarized components

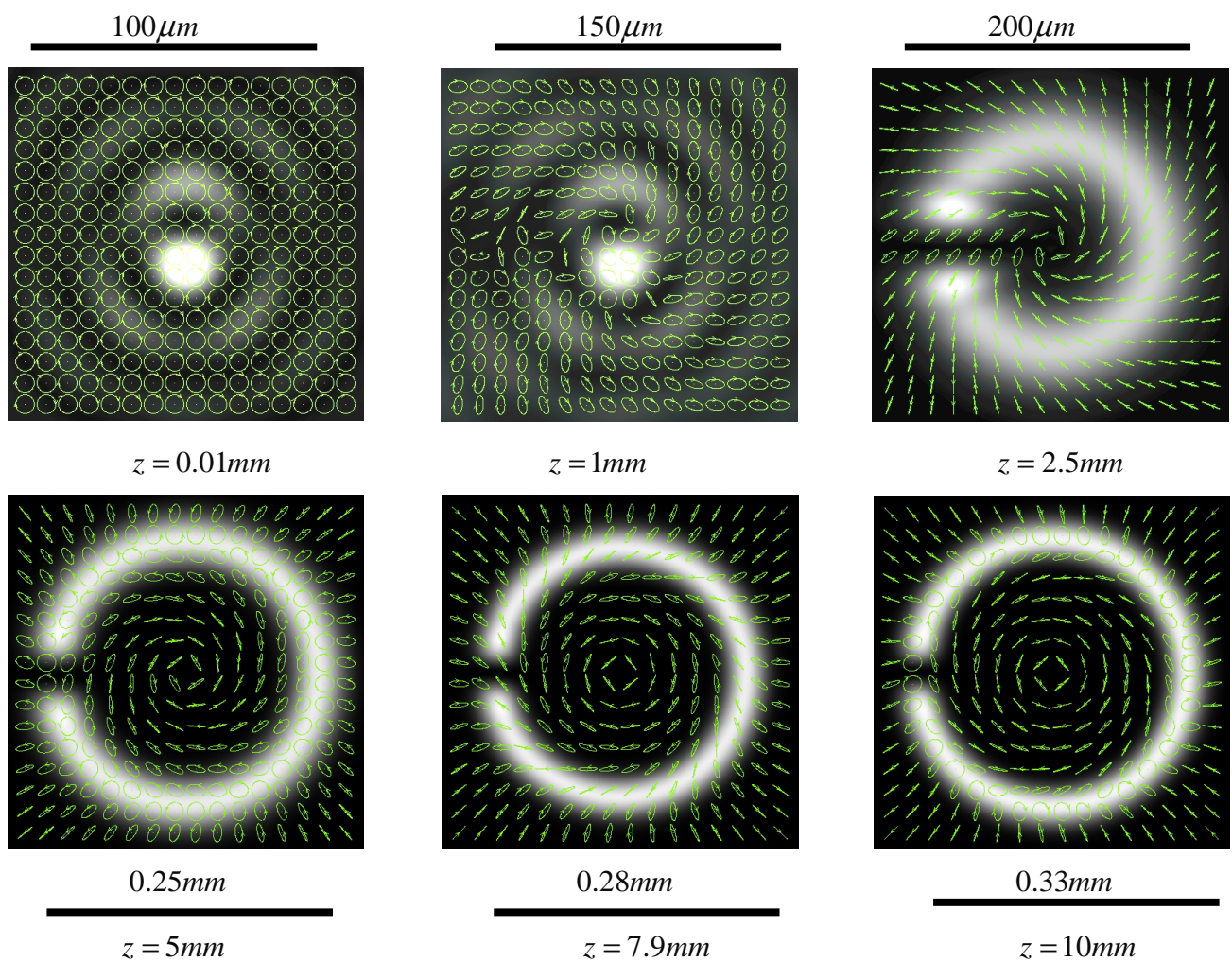

Fig.6 Evolution of the polarization states in the vortex-beam with $K_\perp = i5.5 \cdot 10^5 m^{-1}$, $w_0 = 30 \mu m$

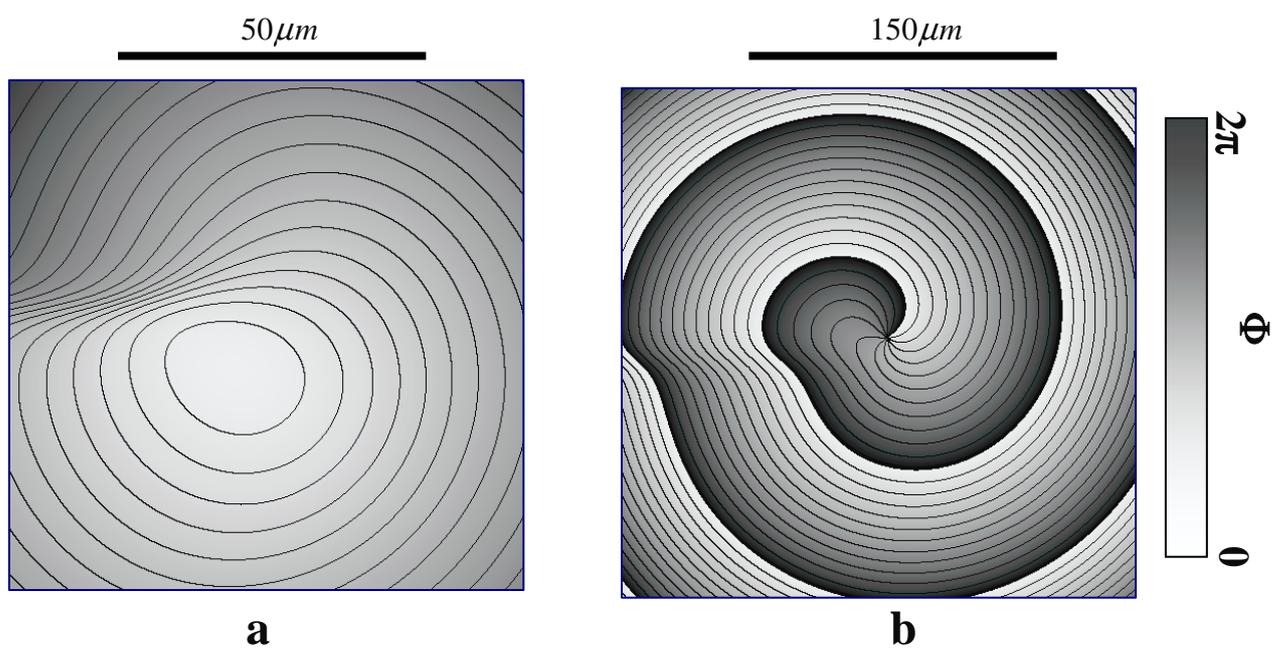

Fig.7 Phase distributions near the beam axis at $z = 10mm$ for (a) $E_+$ and (b) $E_-$ components, $K_\perp = i5.5 \cdot 10^5 m^{-1}$, $w_0 = 30 \mu m$

In order to clarify the physical picture of the above processes let us consider the angular spectrum of the field (21) and (22) at the initial plane $z=0$ ($N_o = N_e = N$, $G_o = G_e = G$). First of all, it should be noted that the field $F_q$ in eq. (24) can be presented as the infinite superposition of the integer order vortices [3]

$$\Psi_q = GN \int_0^{2\pi} e^{iq\phi} \exp\{-K_\perp r \cos(\phi-\varphi)\} d\phi = AGN \sum_{m=-\infty}^{\infty} \int_0^{2\pi} \frac{e^{im\phi}}{q-m} \exp\{-K_\perp r \cos(\phi-\varphi)\} d\phi =$$

$$= 2\pi AGN \sum_{m=-\infty}^{\infty} \frac{e^{im\varphi} I_m(K_\perp r)}{q-m}, \quad (33)$$

where $A = \sin(q\pi) e^{iq\pi}/\pi$, $I_m(x)$ is the modified Bessel function. The spectral function of the field (19) can be written as

$$U_q(\mathbf{K}_p) = \frac{k_o}{2\pi} \iint_\infty E_+(\mathbf{r}) e^{-i(k_x x + k_y y)} d^2\mathbf{r} = \frac{k_o}{2\pi} \int_0^\infty r\,dr \int_0^{2\pi} d\varphi\, E_+(r,\varphi) e^{-iK_p r \cos(\varphi-\phi)}, \quad (34)$$

where $E_+ = -\left\{\Psi_{q-1} + \frac{re^{-i\varphi}}{r_\perp} \Psi_{q-1}\right\}$, $K_p = \sqrt{k_x^2 + k_y^2}$, $\mathbf{r} = (x,y)$, $\mathbf{K}_p = (k_x, k_y)$..

After the forth integration (see 2.12.39.2-3 in [27]) of eq. (34) we come to the expression for the imaginary $K_\perp$ parameter:

$$U_{-1/2}(\mathbf{K}_p) = -NA \frac{k w_0^2}{2} \left\{ \sum_{m=-\infty}^{\infty} \frac{(-i)^m e^{i(m-1)\phi}}{1/2 - m} \left[ e^{i\phi} P_m + \frac{w_0}{r_\perp} Q_m \right] \right\}, \quad (35)$$

$$P_m = \frac{1}{2} \exp\left\{ -\frac{(K_\perp^2 + K_p^2) w_0^2}{4} \right\} I_m\left( -\frac{K_p K_\perp w_0^2}{2} \right), \quad (36)$$

$$Q_m = \frac{(K_\perp K_p w_0)^{2|m|}}{2^{2|m|+1}(|m|!)^2} \sum_{n=0}^{\infty} (-1)^n \frac{\Gamma(n+|m|+3/2)}{n!} \left( \frac{K_p^2 w_0^2}{4} \right)^n {}_2F_1\left(-n, -|m|-n; |m|+1; -\frac{K_p^2}{K_\perp^2}\right), \quad (37)$$

where $\Gamma(v)$ is the Gamma-function, ${}_2F_1$ stands for the hypergeometric function.

Figure 8 illustrates typical angular spectra for two types of the beams. The angular spectrum of the beams with the imaginary $K_\perp = i5 \cdot 10^5 m^{-1}$ (Figure 8a) is characterized by two narrow lines with the extreme values at $K_\perp = i4.5 \cdot 10^5 m^{-1}$ and $K_\perp = i5.5 \cdot 10^5 m^{-1}$. It means that they are shaped by two set of quasi-conical fans of plane waves with the corresponding vertex angles. The beams with the real $K_\perp = 5 \cdot 10^5 m^{-1}$ (Figure 8b) have the smooth spectral line with the maximum value $K_\perp = 0$ associated with the broad spectrum of nearly axial plane waves. It is such nearly axial waves with a very small difference between ordinary and extraordinary refractive indices that provide a weak spin-orbit coupling.

The energy efficiency of the vortex generation in the $E_-$ field component can be estimated by means of the spin-orbit coefficient [28]

$$\eta = \frac{1}{2}\left\{1 - 4\mathrm{Re}\left[\iint_S E_+^o \left(E_+^e\right)^* dS\right] / I_0\right\}, \qquad (37)$$

where $I_0$ is the total intensity of the initial beam at $z=0$, $S$ is the square of the beam cross-section at the arbitrary crystal length.

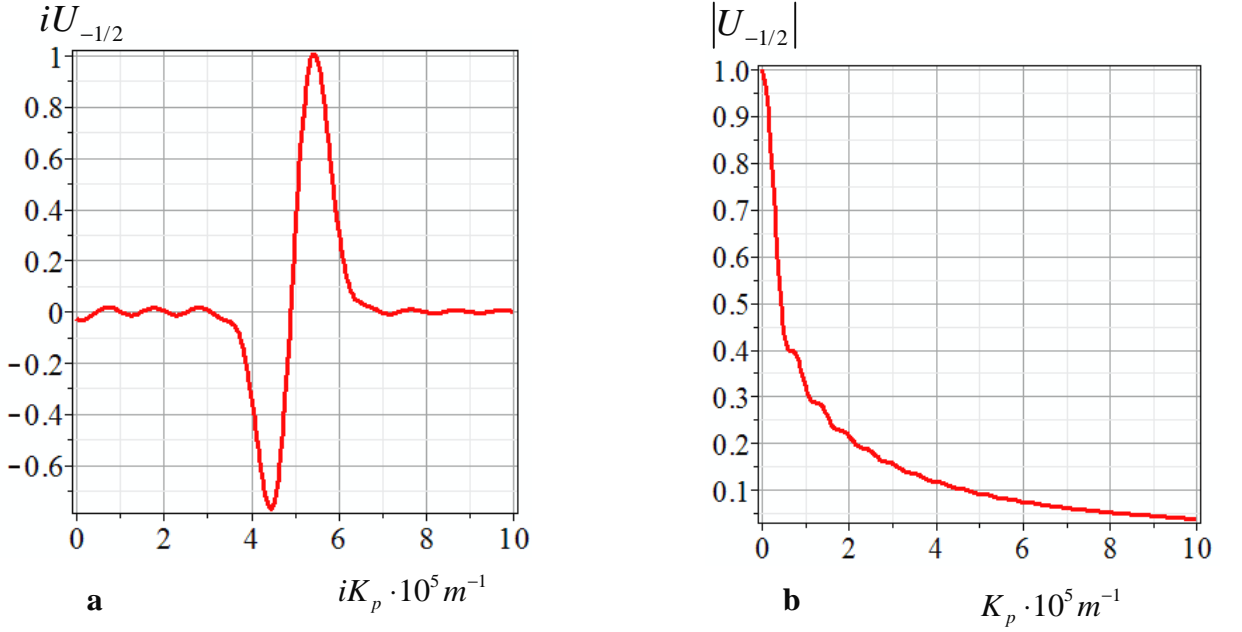

Fig. 8 Angular spectra for the beams with $w_0 = 30\mu m$ and (a) $K_\perp = i5.5 \cdot 10^5 m^{-1}$, (b) $K_\perp = 5.5 \cdot 10^5 m^{-1}$

The beam shaped by a narrow cone of plane waves has a possibility to transform very quickly the energy flux from one circularly polarized component into other than that shaped by a broad spectrum of plane waves. Indeed, the curves $\eta(z)$ with the $K_\perp$ parameter corresponding to the spectral extremes $K_\perp = i5.5 \cdot 10^5 m^{-1}$ and $K_\perp = i4.5 \cdot 10^5 m^{-1}$ shown in Figure 9 have the quasi-periodical forms. Their energy efficiency can reach very high value $\eta = 0.95$. It means that the singly charged optical vortex in the $E_-$ component is generated with the efficiency 95% at the crystal length $z = 4mm$ and $z = 7.5mm$, respectively. (The polarization distributions in Figure 6 were obtained for the crystal lengths corresponding the extreme value $\eta$ and $\eta = 0.5$ of the curve in Figure 9.) However, the energy efficiency in the second type of the beams with the real $K_\perp = 5.5 \cdot 10^5 m^{-1}$ does not exceed 52% at the distance about $z=30mm$. Moreover, this is the maximum value for any arbitrary large crystal length.

## IV. Conclusions

We have analyzed the solutions to the vector paraxial wave equation in the unbounded uniaxial crystal in the form of the transverse electric (TE) and transverse magnetic (TM) mode beams transporting the fractional optical vortices in the circularly polarized components. We have found that both wave types make up two sets of singular beams with the real and imaginary free K parameter. These beam types have different diffraction properties manifesting themselves in processes of the nucleating and breaking down of the fractional optical vortices. We revealed that the TE and TM beams have asymmetric structure in distributing the local elliptic polarization over the beam cross-section. The vector structure of the beams evolves along the beam length. The evolution of the vector structure and the generation of fractional optical vortices are controlled by two processes: the anisotropic diffraction of the beams and the spin-orbit coupling. The fractional optical vortex born in the left-handed circularly polarized component can exist in the form of the holistic structure within small crystal lengths much smaller the Rayleigh. Then it splits into a great number of integer-order optical vortices. At far field, the pair of the singly charged optical vortices with the same signs are created near the beam axis. The beams with the real K parameter cannot form the holistic fractional vortices in both circularly polarized components. They spilt always into a set of integer-order optical vortices. However there is only one singly charged vortex in the circularly polarized component with opposite handedness to the initial one at far field.

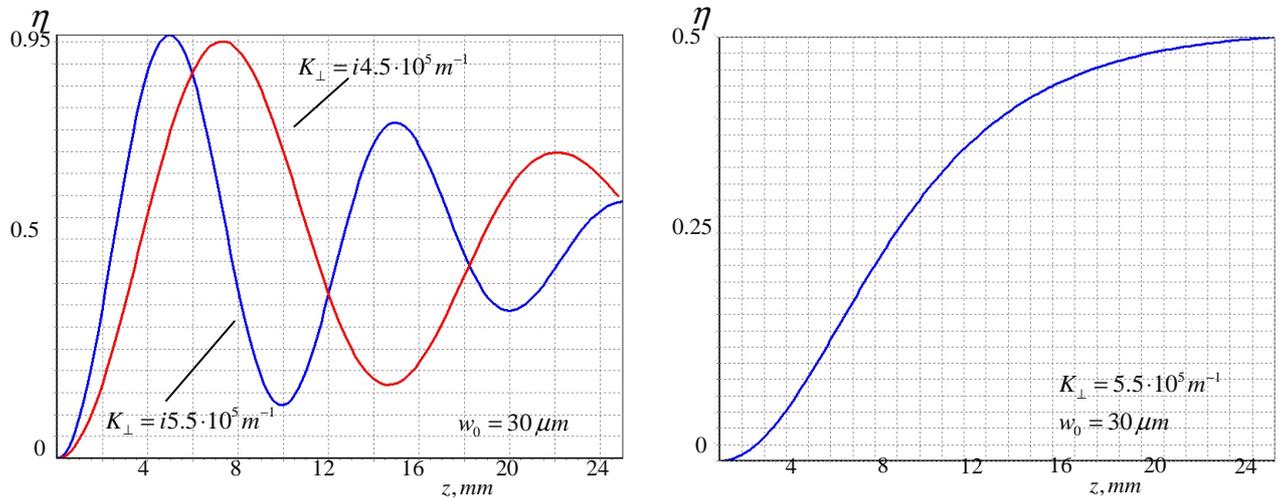

Fig.9 Energy efficiency $\eta$ of the spin-orbit coupling for two types of the vortex-beams

We estimated the energy efficiency of generating the fractional optical vortices and revealed that it can reach 95% in the beams with the imaginary K parameter while the energy efficiency cannot exceed 52% in the beams with the real K parameter. We found that the spin-orbit coupling and, consequently, the energy efficiency is defined by the angular spectrum of the beams. The beam with the real K parameter is characterized by broad spectrum of plane waves propagating at small angles to the crystal optical axis. The beams with the imaginary K parameter are shaped by two conical fans of plane waves. It is this circumstance that define a very high value of the energy efficiency.

**Acknowledgments**

The paper was partially supported by grants 0111U008256 and F41.1/010 of the State Fund for Fundamental Researches of Ukraine.